\documentclass{aa}
\usepackage{graphicx}
\usepackage{txfonts}
\usepackage{natbib}
\bibpunct{(}{)}{;}{a}{}{,} 
\usepackage{bm}

\begin{document}

\title{Structure and dynamics of the class I young stellar object L1489~IRS}

\author{C.~Brinch\inst{1} 
   \and A.~Crapsi\inst{1} 
   \and M.~R.~Hogerheijde\inst{1} 
   \and J.~K.~J{\o}rgensen\inst{2}} 

\date{}

\institute{Leiden Observatory, Leiden University, P.O.~Box 9513, 2300 RA 
		   Leiden, The Netherlands\\
		   \email{brinch@strw.leidenuniv.nl} 
      \and Harvard-Smithsonian Center for Astrophysics, 60 Garden Street, Mail 
	       Stop 42, Cambridge, MA 02138, USA}

\abstract 
{During protostellar collapse, conservation of angular momentum leads to the 
formation of an accretion disc. Little is known observationally about how and 
when the velocity field around the protostar shifts from infall-dominated to 
rotation-dominated.}
{We investigate this transition in the low-mass protostar L1489~IRS, which is 
known to be embedded in a flattened, disc-like structure that shows both infall 
and rotation. We aim to accurately characterise the structure and composition 
of the envelope and its velocity field, and find clues to its nature.}
{We construct a model for L1489~IRS consisting of an flattened envelope and a 
velocity field that can vary from pure infall to pure rotation. We obtain 
best-fit parameters by comparison to 24 molecular transitions from the 
literature, and using a molecular excitation code and a Voronoi optimisation 
algorithm. We test the model against existing millimeter interferometric 
observations, near-infrared scattered light imaging, and $^{12}$CO 
ro-vibrational lines.}
{We find that L1489~IRS is well described by a central stellar mass of 
1.3$\pm$0.4~M$_\odot$ surrounded by a 0.10~M$_\odot$ flattened envelope with 
approximate scale height $h\approx 0.57 R$, inclined at 
${74^\circ}^{+16^\circ}_{-17^\circ}$. The velocity field is strongly dominated 
by rotation, with the velocity vector making an angle of $15^\circ\pm 6^\circ$ 
with the azimuthal direction. Reproducing low-excitation transitions requires 
that the emission and absorption by the starless core $1'$ (8400 AU) east of 
L1489~IRS is included properly, implying that L1489~IRS is located partially 
behind this core.}
{We speculate that L1489~IRS was originally formed closer to the center of this 
core, but has migrated to its current position over the past few times 
$10^5$~yr, consistent with their radial velocity difference of 0.4~km~s$^{-1}$. 
This suggests that L1489~IRS' unusual appearance may be result of its 
migration, and that it would appear as a `normal' embedded protostar if it were 
still surrounded by an extended cloud core. Conversely, we hypothesize that the
inner envelopes of embedded protostars resemble the rotating structure seen 
around L1489~IRS.}


\maketitle

\section{Introduction\label{s:intro}}

The last three decades has provided a detailed understanding of the process of
low-mass star formation through theoretical work and advancements in the 
observational facilities (see for example the review by \citeauthor{andre2000} 
2000; or several reviews in \citeauthor{reipurth2006} 2006). These achievements 
have given us a detailed view on infant stars through their various stages of 
formation. Low-mass stars form out of dark molecular clouds when dense regions 
can collapse under the influence of their own gravity. When sufficient density 
is reached, a protostellar object is formed, still deeply embedded in a 
surrounding envelope. Conservation of angular momentum leads to the formation 
of a disc around the protostar onto which the surrounding dust and gas is 
accreted, although little details are known of how exactly discs grow. As the 
stellar wind starts to clear out the envelope, the star and the disc becomes 
visible in the optical and infrared and the object enters the classical T~Tauri 
stage which then later evolves into a main sequence star 
\citep{shu1977,lizano1989,Adams1988}. Most observed Young Stellar Objects 
(YSOs) are usually classified based on the shape of their spectral energy 
distribution (SED) as either a Class I, II, or III \citep{lada1984}. Class I 
objects are deeply embedded in dense cores, while Class II objects are 
surrounded by actively accreting discs. Class~III objects have little material 
left in a disc, but are still descending to the Main Sequence. Sometimes, 
however, a YSO does not clearly fit into one of these categories. Those objects 
are most likely the ones that can shed light on some of the missing pieces of 
the picture. In this paper we study one such transitional object, L1489~IRS, 
and investigate the structure, dynamics, and composition of its circumstellar 
material.

L1489~IRS (IRAS~04016+2610) is classified as a Class~I object based on its SED 
and visibility at near-infrared wavelengths \citep{myers1987}. Like many 
embedded YSOs, line profiles of dense gas tracers like HCO$^+$ $J$=3--2 and 
4--3 show red-shifted absorption dips usually interpreted as indications of 
inward motions in the envelopes \citep{gregersen2000,mardones1997}. However, 
\citet{hogerheijde2001} shows that the spatially resolved HCO$^+$ $J$=1--0 
emission exhibits a flattened, 2000~AU radius structure dominated by Keplerian 
rotation. In this aspect, L1489~IRS more closely resembles a T~Tauri star with 
a circumstellar disc \citep{koerner1995,guilloteau1998,simon2000}. T~Tauri 
discs, however, are in general much smaller than the disc structure seen in 
L1489~IRS with radii of several hundreds of AU. Scattered light imaging by 
\citet{padgett1999} shows the central stellar object and the presence of a 
slightly flaring dark lane, consistent with the disc-like configuration 
inferred from the HCO$^+$ 1--0 observations. Careful analysis if the 
circumstellar velocity field by~\citet{hogerheijde2001} revealed that infalling 
motions are present at $\sim 10$\% of the Keplerian velocities. 
\citeauthor{hogerheijde2001} hypothesized that L1489~IRS is in a short-lived 
transitional stage between a collapsing envelope (Class~I) and a viscously 
supported, Keplerian disc (Class~II). Observations of ro-vibrational CO 
absorption lines at 4.7$\mu$m showed that the inward motions continue to within
1~AU from the central star \citep{boogert2002}.

In this paper we address a number of questions about L1489~IRS. We construct a 
model for the circumstellar structure that accommodates \emph{all} 
observations, ranging from an extensive set of single-dish molecular line 
measurements to the interferometric observations, the scattered light imaging, 
and the CO ro-vibrational absorption lines. \citet{hogerheijde2001} 
\emph{adopted} a flared disc model with a fixed scale height for the structure 
inspired by the interferometric imaging. In this Paper we choose a description 
for the circumstellar structure that can be smoothly varied from spherical to 
highly flattened, and investigate if the full data set requires a disc-like 
configuration. We also adopt a velocity field that can range from purely 
Keplerian to completely free-fall, or any combination of the two. By 
considering the full data set, stronger constraints can be set on the velocity 
field and the dynamical state of L1489~IRS than possible before. We perform a
rigorous optimisation of the model for L1489~IRS using all available 
single-dish line data, and test the model by comparing the interferometric 
observations, the scattered light imaging, and the CO ro-vibrational absorption 
lines to predictions from the model. Once we have established a satisfactory 
model, also taking into account the immediate cloud environment, we explore the 
nature of L1489~IRS. Does it represent a transitional state between Class~I and 
II? Do all YSOs go through this stage? Or is L1489~IRS is some way special?

The layout of this Paper is as follows. Section~\ref{model} present our data 
set and a detailed overview of the model and fit optimisation procedure. 
Section~\ref{results} describes our best fit model and its reliability, 
including comparison with observations of interferometric observations, 
scattered light imaging, and CO ro-vibrational absorption lines. 
Section~\ref{discus} discusses our results in the light of the nature and 
evolutionary state of L489~IRS and explores the wider implications for our 
understanding of star formation. Section~\ref{sum} concludes the Paper with a 
brief summary.

\section{Observations and model description}\label{model}

\subsection{Single-dish observations}
\begin{figure*}
\begin{center}
\includegraphics[width=18cm]{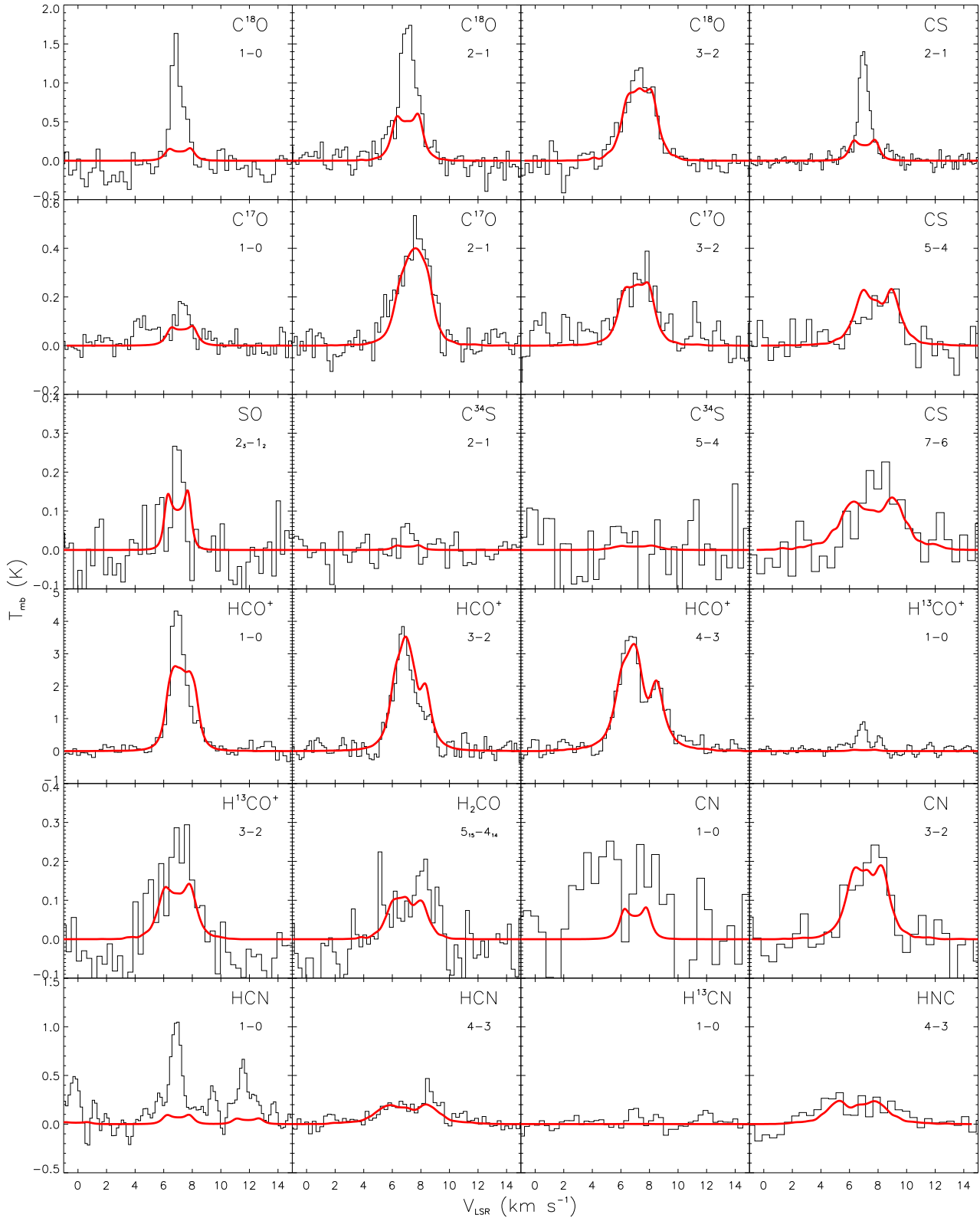}
\caption{The 24 single-dish molecular line spectra used for the model
		 optimisation (histograms). The solid lines show the best-fit
         results for the model of L1489~IRS (See also 
		 Fig.~\ref{core}).
		 }\label{bestfit}
\end{center}
\end{figure*}

The primary data set on L1489~IRS used in this paper was published by
\citet{hogerheijde1997} and \citet{jorgensen2004}, and consists of 24 
transitions among 12 molecular species. Figure~\ref{bestfit} shows all 24 
spectra. Table~\ref{data} lists the transitions, integrated line strengths, 
line widths, and relevant beam sizes of the single-dish telescopes. In all 
cases, line intensities are on the main-beam antenna temperature scale, using 
the appropriate beam efficiencies. The integrated intensities are obtained by 
fitting a Gaussian to the line. In some cases, no lines are visible above the 
noise level, and $3\sigma$ upper limits are given. The signal-to-noise ratio of 
the HNC 4--3 and H$_2$CO $5_{15}$--$4_{14}$ spectra was insufficient for a 
proper Gaussian fit; instead the spectra are simply integrated from $-4$ to 
$+4$~km~s$^{-1}$ with respect to the systemic velocity of $+7.2$~km~s$^{-1}$. 
In addition to these molecular line data, we also use the total mass derived 
from the 850 $\mu$m continuum observations by JCMT/SCUBA 
\citep{hogerheijde2000}.

\begin{table}
\caption{Single-dish Molecular Line Data Set}\label{data}
\begin{tabular}{l l c c c}
\hline \hline
         &            & $\int T_{\rm mb}$d$v$& FWHM & Beam  \\
Molecule & Transition & (K km s$^{-1}$) &(km s$^{-1}$)& ($''$)\\ 
\hline 
C$^{18}$O		& 1--0 & 1.8 $\pm$ 0.04	& 1.1 & 34\\
				& 2--1 & 2.8 $\pm$ 0.1	& 1.6 & 23\\
				& 3--2 & 3.4 $\pm$ 0.1	& 2.5 & 15\\
C$^{17}$O		& 1--0 & 0.5$^a$ $\pm$ 0.04	  & 2.9 & 22\\
 				& 2--1 & 1.2$^a$ $\pm$ 0.03   & 2.6 & 11\\
				& 3--2 & 0.7 $\pm$ 0.1	& 2.7 & 15\\
CS				& 2--1 & 1.2 $\pm$ 0.02	& 0.9 & 38\\
				& 5--4 & 0.6 $\pm$ 0.04	& 0.8 & 22\\
				& 7--6 & 0.7 $\pm$ 0.1  & 3.6 & 15\\	
C$^{34}$S		& 2--1 &$<$0.3$^b$      & -- & 39\\
				& 5--4 &$<$0.3$^b$      & -- & 21\\ 
HCO$^+$			& 1--0 & 6.7 $\pm$ 0.2	& 2.1	& 28\\
				& 3--2 & 6.9 $\pm$ 0.2	& 2.2	&19\\
				& 4--3 & 10.0 $\pm$ 0.3 & 2.5	&14\\
H$^{13}$CO$^+$	& 1--0 & 0.8 $\pm$ 0.1	& 0.8 & 43\\
				& 3--2 & 0.8 $\pm$ 0.1	& 1.8 & 19\\
HCN				& 1--0 & 2.4$^a$ $\pm$ 0.1   & 2.3 & 43\\
				& 4--3 & 1.3 $\pm$ 0.1	& 5.8 & 14\\
H$^{13}$CN  	& 1--0 &$<$0.5$^b$      & -- & 44\\
CN 				& 1$_{023}$--0$_{012}$ 	& 0.63 $\pm$ 0.12 & -- & 33\\
				& 3--2 & 0.67$^a$ $\pm$ 0.09		& 3.9 & 15\\
HNC				& 4--3 & 1.5 $\pm$ 0.3& 6.4  & 14\\
H$_2$CO			& 5$_{15}$--4$_{14}$ & 0.61 $\pm$ 0.07& 4.0 & 14\\
SO				& 2$_3$--1$_2$ 	& 0.3 $\pm$ 0.03& 0.8 & 38\\
\hline
\end{tabular}

$^a$ Intensity integrated over multiple hyperfine components.\\
$^b$ No line detected. 3$\sigma$ upper limit, based on an assumed line width 
of 1.5 kms$^{-1}$.\\
The error bars on the intensity does not include the 20\% calibration error.
\end{table}

Apart from the single dish data which we use in the model optimisation, we 
compare predictions by the model to other previously published observations: a 
HCO$^+$ $J$=1--0 interferometer map from the BIMA and OVRO arrays 
\citep{hogerheijde1998}, CO ro-vibrational absorption spectra from the Keck 
Telescope\citep{boogert2002}, and the near-infrared scattered light imaging 
from HST/NICMOS \citep{padgett1999}.

\subsection{The model}

Previous studies of L1489~IRS clearly indicate that an axi-symmetric 
description of its circumstellar structure is required. In the following 
subsections we construct a description of the density $n(r,\theta)$, gas 
temperature $T(r,\theta)$, and velocity field 
$\vec v(r,\theta)$=$(v_R,v_z,v_\phi)$. Throughout we attempt to keep the number 
of free parameters at a minimum. In the end we arrive at four free parameters, 
in addition to the eight molecular abundances which we fit but assume constant 
throughout the source, and seven parameters that we hold fixed 
(Table~\ref{parameter}).

\subsubsection{Density}

We adopt an axi-symmetric description of the gas density $n(R,z)$ consistent 
with the spherical model from \citep{jorgensen2002}. These authors deduce a 
total mass of 0.097~M$_\odot$ and a density following a radial power Law with 
slope $-1.8$ between radii of 7.8 and 9360~AU. We truncate this model at the 
observed outer radius of L1489~IRS of 2000~AU, but keep the power-law slope and 
mass conserved. Instead of a simple radial power law, $n\propto r^{-p}$ with 
$p=1.8$, we adopt a Plummer-like profile, $n\propto [1+(r/r_0)^2]^{-p/2}$ with 
$r_0$=4.0~AU. This description keeps the density finite at all radii, but since 
$r_0$ is much smaller than the scales of interest here, the resulting density 
distribution is identical to that used by \citep{jorgensen2002}.

From this spherically symmetric density distribution we construct an
axi-symmetric, flattened configuration by multiplying by a factor
$\sin^f\theta$, where $f$ can take any value $\ge 0$ \citep[see][where this 
approach was used for modelling starless cores]{stamatellos2004}. The adopted 
density distribution now becomes,
\begin{eqnarray}
n(r,\theta) = 
	n_0 \Bigl(1+\bigl({r\over{r_0}}\bigl)^2\Bigr)^{-p/2} \sin^f\theta.
\label{density}
\end{eqnarray} 

\begin{figure}
\begin{center}
\includegraphics[width=8cm]{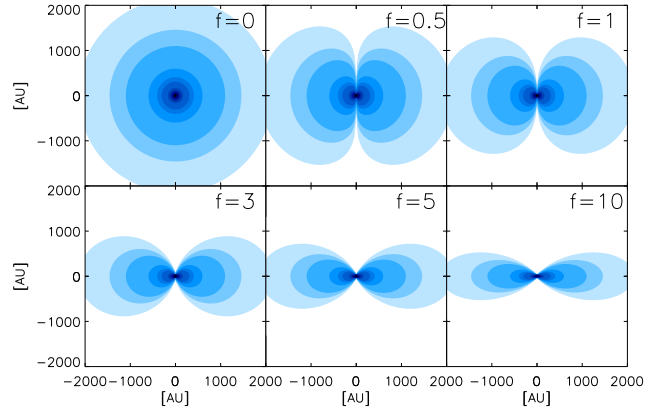}
\end{center}
\caption{Progression of flattening of the adopted density structure as 
	     $f$ is increased from 0 (purely spherical) to 10 in Eq. 
		 (\ref{density}).
		 }\label{flatness}
\end{figure}

For $f=0$ this reduces to a spherically symmetric structure, while for $f>10$ 
the resulting profiles becomes largely indistinguishable as they approach a 
step function (Fig.~\ref{flatness}). The mass contained in the structure is 
kept constant at 0.097~M$_\odot$ by adjusting $n_0$ as $f$ is varied. The only 
free parameter in the density description is the flattening parameter $f$.

\subsubsection{Temperature}

The temperature of the gas and the dust (which we assume to be identical) 
in the circumstellar structure of L1489~IRS depends on the stellar luminosity 
which is $\sim 3.7$~L$_\odot$~\citep{kenyon1993a} and the infrared radiative 
transfer through the structure. Since most of the circumstellar material is 
optically thin to far-infrared radiation, the deviations introduced by the 
flattening on the temperature structure are minor. Furthermore, the line 
excitation does not depend strongly on small temperature differences. A 
spherically symmetric description of the temperature therefore suffices. Using 
the continuum radiation transfer code DUSTY~\citep{nenkova1999} and the density 
structure of Eq.~(\ref{density}) with $p=1.8$ and $f=0$, we find that the 
temperature is well described by,
\begin{eqnarray}
T(r)=19.42\, {\rm K} \, \Bigl({r\over{1000\,{\rm AU}}}\Bigr)^{-0.35}.
\end{eqnarray}

There are no free parameters in this description of the temperature.

\subsubsection{Velocity field}

The velocity field is parameterized by a central, stellar mass $M_\star$ and an 
angle $\alpha$ in such a way that,
\begin{eqnarray}
v_r &= -\sqrt{2}\,\sqrt{ {GM_\star}\over{r}} \sin\alpha,\\
v_\phi &= \sqrt{ {GM_\star}\over{r}} \cos\alpha.\label{velocity}
\end{eqnarray}

For $\alpha=0$ this reduces to pure Keplerian rotation around a mass $M_\star$ 
without any inward motions; for $\alpha={\pi\over 2}$ the velocity field is 
that of free fall to a mass $M_\star$. Intermediate values of $\alpha$ produce 
a velocity field where material spirals inward. The implicit assumption in this 
description is that both components of the velocity field vary inversely 
proportional with $\sqrt{r}$. Note that $\alpha$ should not be 
confused with the geometric angle determining the direction of the flow lines.

In this description there are two free parameters, the stellar mass $M_\star$ 
and the angle $\alpha$ which is kept constant with radius. In addition to this 
ordered velocity field, we add a turbulent velocity field with FWHM 
0.2~km~s$^{-1}$.

\subsection{Molecular excitation and line radiative transfer}

\begin{figure}
\begin{center}
\includegraphics[width=8cm]{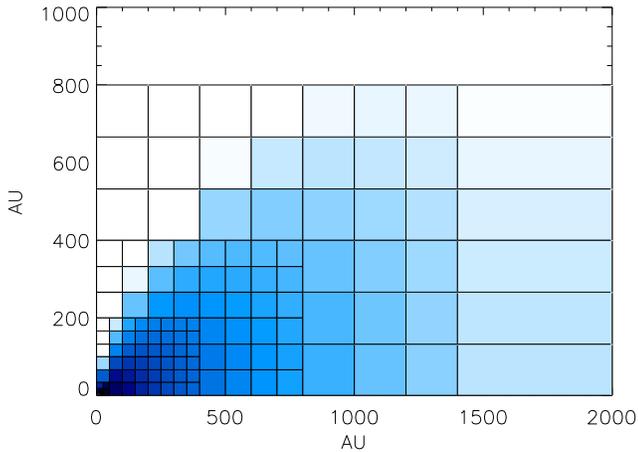}
\end{center}
\caption{Layout of the grid cells for a model with $f=3.8$.}\label{modelplot}
\end{figure}

The excitation of the molecules and the line radiative transfer is calculated 
using the Accelerated Monte Carlo code RATRAN \citep{hogerheijde2000a}. 
Collisional excitation rates are taken from the Leiden Atomic and Molecular 
Database LAMDA \citep{schoier2005}. We lay out the model onto three nested 
$8\times 6$ grids (Fig. \ref{modelplot}). The innermost grid cell is subdivided 
four times, so that the innermost cell is resolved down to 4~AU. All 
properties are calculated as cell averages, by numerically integration over the 
cell and divide by its volume. To reduce 
computing time, cells with H$_2$ densities below $10^3$~cm$^{-3}$ are dropped. 
Such cells do not contribute significantly to the line emission or absorption. 
Dust continuum emission is included through a standard gas-to-dust ratio of 
100:1 and dust emissivity from \citet{ossenkopf1994} for thin ice mantles which 
has been accreted and coagulated for about 10$^5$ years.

Synthetic observations are created from the molecular excitation by performing 
ray-tracing after placing the object at a distance of 140~pc and an inclination 
$i$ (a free parameter). The resulting spectra are convolved with the 
appropriate Gaussian beams. Figure~\ref{bestfit} shows the best-fit model 
spectra (the best fit is discussed in section~\ref{results}).

\subsection{Modelling the neighbouring cloud core}

During the optimisation of the fit (\S\ref{opt}) it became obvious that several 
lines, and especially those of lower-lying rotational transitions taken in 
large beams, were contaminated by emission with small line width. This emission 
component is especially clear in the C$^{18}$O 1--0 and 2--1 lines, the CS 2--1 
line, the HCN 1--0 line, and, to some extent, the HCO$^+$ 1--0 line 
(Fig.~\ref{bestfit}). The emission has a $V_{\rm LSR}$ of 6.8~km~s$^{-1}$, 
slightly lower than that of L1489~IRS of 7.2~km~s$^{-1}$. Cold fore- or 
background gas with small turbulent velocity is the likely cause for this 
component. The 850 $\mu$m SCUBA map from \citet{hogerheijde2000} reveals that 
L1489~IRS sits at the edge of an extended, probably starless, cloud core with a 
radius of $60''$ (8400 AU). Cold gas in this core therefore contributes to the 
low-$J$ emission lines, and especially in spectra taken with large beams.

We construct a simple description for the neighbouring core, so that we can 
take its emission into account in our optimisation of the model for L1489~IRS, 
as well as its absorption if this source is located behind the core. We 
approximate the core as spherical with a radius of $60''$, which is roughly the 
distance of L1489~IRS to its centre. We assume that it is isothermal at 10~K 
and that it has abundances typical for starless cores \citep{jorgensen2004}. 
For the species which does not show any cloud core emission, the abundances are 
unconstrained and we just set the abundances sufficiently low. In the case of 
CO we use an abundance of $5\times10^{-5}$. The CS abundance is set to 
$2\times 10^{-9}$, and the HCO$^+$ and HCN abundances are $27\times10^{-9}$ and 
$4\times10^{-9}$ respectively. We derive its density distribution by fitting 
the 850~$\mu$m emission from \citet{hogerheijde2000}. We find an adequate fit 
for a radial power-law with slope $-2$ and a density of 
$4\times 10^6$ cm$^{-3}$ at $r=1000$~AU resulting in a cloud mass of 
2.9~M$_\odot$. This is consistent with the drop off in density found in many 
starless cores on scales ($r > 1000$~AU) that are relevant to us 
\citep{andre1996}. Because it falls outside even our largest beam on L1489~IRS 
we do not investigate if the density in the neighbouring core levels off at the
center, as is seen for many starless cores. The relative smoothness of the 
850~$\mu$m emission suggest that this is the case, however.

Using RATRAN we calculate the expected emission and the optical depth of each of 
the observed transitions. In our model optimisation procedure (see below), the 
emission from L1489~IRS and the neighbouring core are added on a 
channel-by-channel basis, with the appropriate spatial offset for the core. We 
find that we can only make a fit that is reasonable if L1489~IRS is located 
behind the core; we need both the emission and the opacity of the cloud. This 
is taken into account by first attenuating the emission from L1489~IRS by the 
core's opacity, again on a channel-by-channel basis, and subsequently 
adding the core's emission in each channel, followed by beam convolution.

In this section we derived only an approximate model for the neighbouring core. 
Its effects are taken into account in the model spectra, but the description of 
the core is not accurate enough to include in the model optimisation. This 
would require a much more detailed analysis than possible here. In the 
procedure outlined in the next section, we therefore mask out those regions in 
the spectra strongly affected by the emission and absorption of the core.

\subsection{Optimising the fit}\label{opt}

Our model has four free parameters: the inclination $i$, the flatness parameter 
$f$, the stellar mass $M_\star$, and the angle of the velocity field $\alpha$. 
In addition, the abundances of the molecules are unknown. All other parameters 
are held fixed. Table~\ref{parameter} lists the parameters.

Considering the size of the parameter space and the time it takes to calculate 
a single spectrum\footnote{Depending on the species and the
	optical thickness, we can calculate a spectrum in between five
	minutes and half of an hour, on a state-of-the-art desktop
	processor.} 
the task of finding the parameter vector resulting in the best fit is 
non-trivial. This is further complicated by the degeneracy of the model results 
to different parameters. For example, increasing the abundance can have the 
same effect on the line intensity as increasing the inclination or the 
flatness, but these will have very different effects on the line profile shape.

Instead of calculating all possible models in the allowed parameter space, we 
use Voronoi tessellation of the parameter 
cube~\citep[see e.g.][for details on Voronoi tessellation]{kiang1966}. A random 
set of $n$ points $p_n$ in the parameter cube is picked and model spectra are 
calculated for each of these. Then the parameter cube is divided into Voronoi 
cells, defined as the volume around a point $p_i$ in the parameter cube 
containing all points $q$ closer to $p_i$ than to any other of the points $p_n$ 
($n\neq i$). The parameter cube is scaled in arbitrary units, so that the 
allowed parameter ranges falls between 0 and 1. On this dimensionless unit cube 
a simple metric in $d$ dimensions is used to define the cells,
\begin{eqnarray}\label{distance}
s^2 = \sum_{i=1}^d \left ( q_i - p_i\right )^2,
\end{eqnarray}
assuming that the solution depend linearly on all parameters. This assumption 
is not true especially for large values of $s$, but because we have no 
knowledge of the geometry of the parameter space, we use the simplest possible 
measure. In order to minimise the effect this has on our final solution we can 
increase the initial sample rate so that the average distance between the 
points becomes smaller. After one or two iterations, the volume of each cell is 
small enough so that the assumption of linear dependence is good. By scaling 
the parameters to the same range we make sure that each parameter is weighted 
equally in the distance measure.

The cell which contains the point $p_i$ resulting in the best fit is chosen, 
and a new set of random points are picked within this cell, and the procedure 
is iterated until sufficient convergence has been achieved. This method is only 
guaranteed to reach the true best fit if only one global minimum exist and if 
there are no (or few) local minima. To check whether we find the true optimum, 
we make several runs, with different randomly distribution initial points. We 
find that we always reach the same minimum, and conclude that local minima are 
few and not very deep.

For every calculated model spectrum, the fitness is evaluated by regriding the 
model spectrum to the channel width of the corresponding observed spectrum, 
centering it on the LSR velocity of 7.2 kms$^{-1}$, and calculating the 
$\chi ^2$ between the model and the observed spectrum,
\begin{eqnarray}
\chi ^2 = \frac{1}{M}\sum_m \frac{1}{N_m}\sum_n \frac{(I(n)_{obs}-I(n)_{model})^2}{\sigma ^2},
\end{eqnarray}
where $M$ is the number of spectra and $N$ is the number of velocity  channels 
in the $m$'th spectrum. This way we give an equal weight to all spectra even 
though the number of channels vary in each spectrum. Those channels affected by 
the neighbouring core are not included in the $\chi^2$ measure. Every spectra 
has a fixed passband of 14 kms$^{-1}$ so that an equal amount of baseline is 
included for each spectrum.

Using this method, with a set of 24 random points per iteration, we converge on 
an optimal solution after four to five iterations, corresponding to 10 to 12 
days of CPU time. For practical reasons we initially chose only to consider the 
most structured lines (CO, HCO$^+$ and CS), lowering the computational time to 
about a single day and getting a quick but rough handle on the initial 
parameter cube. We then included the other lines to obtain the overall best 
solution.

\begin{figure}
\includegraphics[width=8.5cm]{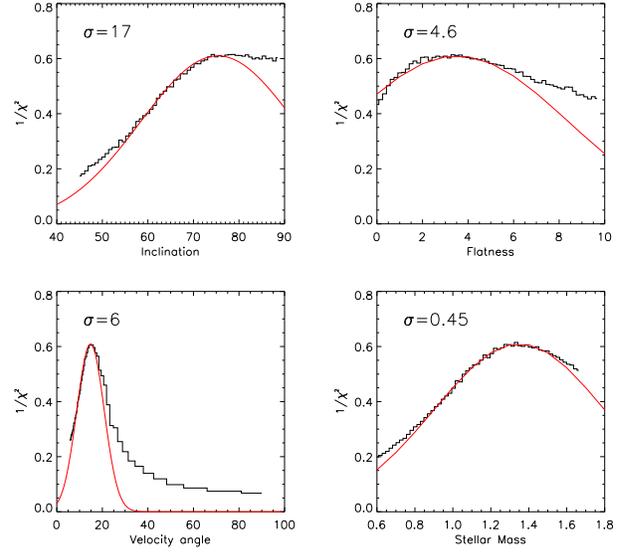}
\caption{The histograms show the $\chi^2$ distributions of models around the
		 best fit position where only one parameter is varied at each time. 
		 A Gaussian, centered on the best fit in each panel, is fitted to the 
		 distributions. The dispersion of the gaussians is given in each panel. 
		 }\label{conv}
\end{figure}

\subsection{Error estimates}

Getting a handle on the uncertainties in the obtained parameter values is a 
difficult matter due to the size and complexity of the parameter space. As 
mentioned above, we have no knowledge of the overall geometry of the parameter 
space and given the long computation time of the optimisation algorithm, we 
cannot make a correlation analysis of each pair of parameter and neither can we 
make $\chi ^2$ surfaces. Still, it is very important to get an estimate on the 
stability and reliability of our solution.

A simple error analysis is done for the four model dependent parameters, the 
flatness, the velocity angle, the stellar mass, and the inclination, by fixing 
three of the parameters at their best fit values, and calculating models in 
which the fourth parameter is gradually increased from its lower boundary to 
the upper boundary. Histograms of the resulting (inverse) $\chi ^2$ values is
shown in Fig.~\ref{conv}. The $\chi ^2$ values are approximately normally 
distributed, with the main discrepancy in the high values of the inclination, 
velocity field, and the flatness. This relates to the non-linear nature of the 
trigonometric functions associated with these three parameters. 

A Gaussian has been fitted to each of the histograms in Fig~\ref{conv}. The
centre of the Gaussian is fixed on the best fit value and the hight is fixed by 
the $\chi ^2$ value of the best fit so that only the variance, $\sigma^2$, is 
free. Reasonable fits are achieved for each parameter with the $\sigma$ value 
given in each panel. These values are taken to be a rough estimate of the
magnitude of the error in each of the four parameters. For the inclination and 
flatness, where the error is greater than the allowed parameter range, the 
error is of course determined by the physical constrains on the parameter value 
(e.g., the inclination cannot be greater than $90^\circ$). Note that the error 
bars are typically smaller than the explored range in each parameter by a 
factor of 2--10. 

With this kind of one dimensional error analysis we do not take into account 
the fact that the parameters are likely to be highly correlated. A few 
exploratory calculations, where one parameter was held fixed at its best fit
value while the other three where randomly pertubed around their best fit 
values, indicated that there indeed exist a strong correlation between the 
parameters. Indeed, a degeneracy exists between the central mass and the 
inclination, which again is degenerate with the flattening. Only a full 
parameter space study can fully disentangle this and is beyond the scope of
this paper.

Because the abundance parameter mainly serves to scale the intensity in every
channel of the spectrum, and does not change the shape of the profile much, 
this kind of error analysis is of little use. For any combination of the four 
free parameters that reproduces the observations, a corresponding abundance is 
found from the optically thin isotopic lines. These abundances are relatively 
insensitive to the exact geometry because of their optically thin nature. 
Therefore we assume that the error in the abundance values obtained here is 
entirely dominated by the 20\% calibration error of the observed spectra.

Throughout this work we have assumed constant abundance for all the molecular 
species. In reality, abundances will depend on the chemistry and molecules will 
freeze out below a certain temperature. This gives rise to a drop in the 
abundances at a certain radius and it will affect, to some extent, the shape of 
the profiles but more prominently, the line ratios. Specifically, by removing 
low temperature material from the gas phase, low excitation lines become 
relatively weaker. Our model does not suffer from the problem of over-producing 
the low $J$ lines, except for the case of HCO$^+$; a more complex abundance 
model would likely provide a better fit to the $J=$ 1--0 and 3--2 lines. 
However, this would require a careful chemical analysis which is beyond the 
scope of this paper. A few tests showed that letting CO freeze out at 20 K does 
not change the best fit parameters significantly, except for the abundance 
which will then have to be re-optimised.

\section{Results}\label{results}

\begin{figure*}
\begin{center}
\includegraphics[width=18cm]{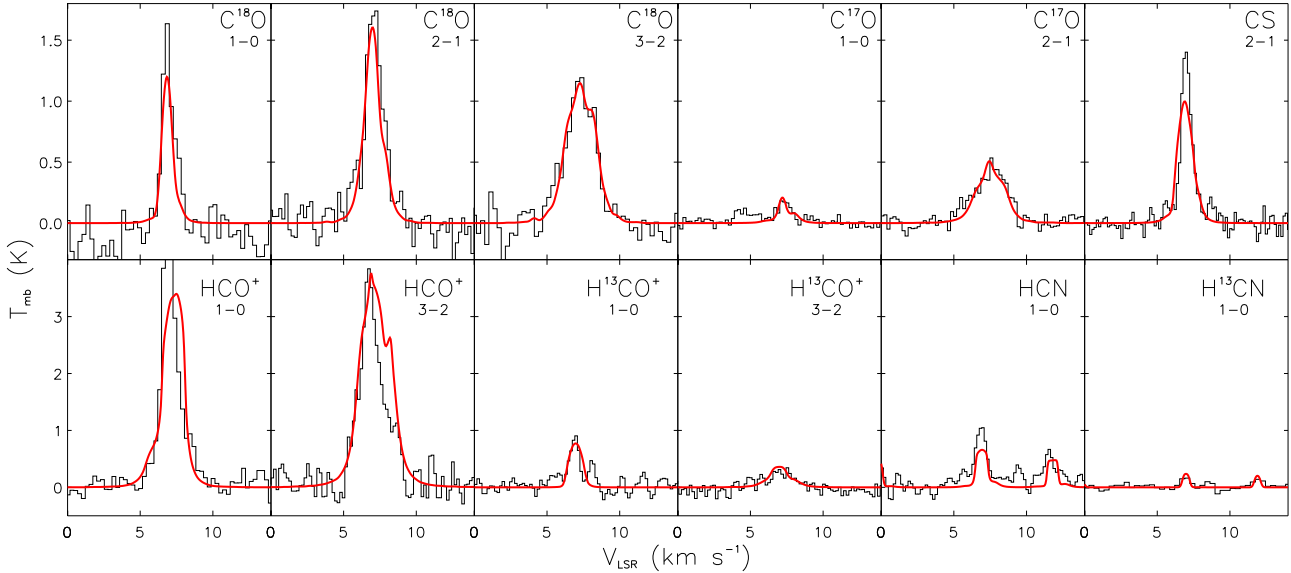}
\caption{The combined emission from the L1489~IRS model and the neighbour 
		 cloud core. Only spectra which are affected significantly by the 
		 cloud core emission are shown here.
		 }\label{core}
\end{center}
\end{figure*}

Figure~\ref{bestfit} compares the data to the synthetic spectra based on the 
best fit model obtained with the optimisation procedure described above. The 
results for the combined emission of L1489~IRS and the neighbour core is shown 
in Fig.~\ref{core}. Because the emission from the core only contributes to the 
low $J$ lines,  this figure only shows the species in which the combined 
spectrum show any difference from the L1489~IRS spectrum alone. For all species 
not shown in Fig.~\ref{core}, the combined spectra is indistinguishable from 
the one shown in Fig.~\ref{bestfit}.
Table~\ref{parameter} lists the parameters of the best fit.

\begin{table}
\caption{Best Fit Parameters}\label{parameter}
\begin{center}
\begin{tabular}{l c}
\hline \hline
Free Parameter & Value \\
\hline 
Inclination of disc, $i$& 74$^\circ$ \\
Flatness of disc, $\sin^f(\theta)$ & 3.8 \\
Central mass, M$_*$ & 1.35 M$_\odot$ \\ 
Velocity angle, $\alpha$ & 15$^\circ$ \\
\\
Abundance of CO$^a$ & 	$3.5\times10^{-5}$\\
Abundance of CS&   		$0.5\times10^{-9}$\\
Abundance of SO &  		$2.0\times10^{-9}$\\
Abundance of HCO$^+$ & 	$1.9\times10^{-9}$\\
Abundance of HCN & 		$0.2\times10^{-9}$\\
Abundance of HNC & 		$0.2\times10^{-9}$\\
Abundance of CN &  		$0.2\times10^{-9}$\\
Abundance of H$_2$CO & 	$0.7\times10^{-9}$\\  
\hline
Fixed parameters & \\
\hline
Disc mass& 0.097 $M_\odot$\\
Temperature slope & $-0.35$\\
Temperature at 1000~AU & 19.42~K\\ 
Density slope & $-1.8$\\
Turbulent velocity, FWHM & 0.2 km s$^{-1}$\\
Outer radius & 2000 AU\\
Distance & 140 pc\\
\hline
\end{tabular}
\end{center}
$^a$ The main isotopic abundance has been derived from the C$^{18}$O
abundance using a $^{16}$O/$^{18}$O ratio of 540~\citep{wilson1994}.
\end{table}

The inclination angle of 74$^\circ$ falls within the range of 60$^\circ$ to 
90$^\circ$ which is inferred from the scattered light image of 
\citet{padgett1999} and the modelling of the infrared spectral energy 
distribution \citet{kenyon1993}. Section \ref{nir} shows that this inclination 
and the flattening parameter $f=3.8$ reproduce the scattered light image, 
including the detectability of the central star. The resulting density 
distribution can be well approximated by a disk with a vertical density 
distribution $\propto e^{-z^2/2h^2}$ and a scale height of $h\approx 0.57 R$.
The maximum deviation of this approximation is only 4\%, up to an angle of 
60$^\circ$ above the midplane.
\citet{hogerheijde2001} described L1489~IRS with flared disc with an adopted 
density scale height of $h = 0.5 R$, so our best fit model shows that the 
structure might in fact be flatter than previously assumed. However,
\citet{hogerheijde2001} used an inclination of $90^\circ$ for the flared disc 
(a flat disc at $i=60^\circ$ was also tried), so the projected column density 
distribution are quite similar in both cases.

The best fitting velocity vector makes an angle of only 15$^\circ$ with respect 
to the azimuthal direction, consistent with \citet{hogerheijde2001} who shows 
that rotation is the primary component in the velocity field. However, our 
central stellar mass of 1.35~$M_\odot$ is considerably higher than the 
0.65~$M_\odot$ derived by \citeauthor{hogerheijde2001}. Part of this is due to 
different definitions of the velocity field. \citeauthor{hogerheijde2001} 
reports a mass expected for Keplerian motion based on the azimuthal component 
of the velocity field alone, while the mass derived here results in 
$\cos\alpha$ times the Keplerian velocity (Eq. \ref{velocity}). In addition, a 
different inclination is found. These two factors together would give a mass of 
1.2~M$_\odot$ for our model. This mass is still 80\% higher than that from 
\citeauthor{hogerheijde2001} but our best fit has a $\chi^2$ value of 1.69. 
That is nearly half that of the best model of \citeauthor{hogerheijde2001} of 
$\sim$3. We ascribe this difference to a more thorough search of the parameter 
space.

The best fitting abundances are consistent with the abundances obtained for 
L1489~IRS in~\citep{jorgensen2004} to within a factor of 2--3 although the 
values obtained here are all higher. This may be due to higher line opacities 
in our model as a result of the different adopted velocity fields; 
\citeauthor{jorgensen2004} do not include a systematic velocity field and only 
a single turbulent line width. Consistent with this previous work, we find no 
evidence for depletion of CO, in accord with the relatively high temperatures
exceeding the 20~K evaporation temperature of CO throughout most of the disc. 
Another interesting finding is the CN/HCN abundance ratio of 1.0, which is more 
reminiscent of dark clouds than of circumstellar discs, suggesting that 
chemically, L1489~IRS is close to its original cloud core and that 
photo-dissociation does not play a major role yet \citep{thi2004}.

\subsection{Quality of the best fit}

The overall correspondence of our model to the data is good and most of the 
spectra are well reproduced. The line widths of C$^{18}$O and C$^{17}$O are 
well reproduced, and the C$^{17}$O 2--1 and 3--2 lines are found to exclusively 
trace L1489~IRS and be uncontaminated by the neighbouring core. All three 
C$^{18}$O and the C$^{17}$O 1--0 line have narrow line peaks that originate in 
the neighbour core; some C$^{18}$O $J=$ 1--0 emission that is not reproduced can 
either be caused by additional material along the line of sight, or be due to 
the approximate nature of our description of the core.

The sulfur bearing lines are not very intense and show little structure. Again 
we see a narrow peak in the CS $J=$ 2--1 almost entirely accounted for by the 
neighbouring core and perhaps also in the SO line. The non-detection of 
C$^{34}$S 2--1 places an upper limit on the CS abundance in the neighbouring 
cloud of $2\times10^{-9}$. The CS 7--6 line is poorly fit, but the observed 
spectrum has low signal-to-noise.

The HCO$^+$ lines are the strongest among our sample and show most structure in 
their profiles. It was important to be able to reproduce the double peak in the 
HCO$^+$ $J=$ 4--3 line because this feature is a very clear tracer of the 
velocity structure in L1489~IRS. Our model is able to reproduce this feature. 
However, we do not observe the double peak in the HCO$^+$ $J=$ 3--2 line but 
only a slight asymmetry. This provided a major constraint on the velocity 
field. The neighbour core does not contribute in the 4--3 line but it makes up 
for almost all of the excess emission seen in the 1--0 and 3--2 lines of 
HCO$^+$.

Of the nitrogen bearing species, HCN $J=$ 1--0 with its three hyperfine 
components shows very narrow lines, which is well reproduced by the neighbour 
core model which dominates the emission. The HCN $J=$ 4--3 line is much broader 
and uncontaminated by the neighbouring core. A narrow peak in the spectrum at 
8~km~s$^{-1}$ cannot be due to the neighbour core and we assume it is noise. 
The HNC 4--3 line has a rather low signal to noise ratio and can only provide a 
reliable estimate within a factor of a few. CN $J=$ 1--0 also has a low signal 
to noise ratio, but in the case of CN, the abundance is well constrained by the 
$J=$ 3--2 line.
\begin{figure}
\begin{center}
\includegraphics[width=8.8cm]{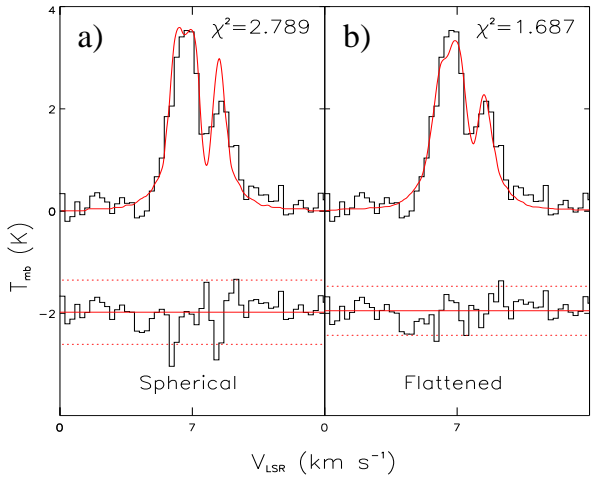}
\caption{a) Best possible fit with a spherical model to the HCO$^+$ $J=$ 4--3 
		 line. b) The best fit to the same line where the flattening $f$ and 
		 the inclination $i$ are kept as free parameters. Below, the residuals 
		 are shown with the mean and two standard deviations indicated.
		 }\label{duo}
\end{center}
\end{figure}

All spectra discussed above have been obtained with single-dish telescopes that 
do not resolve the 2000~AU radius source. One can wonder if it is justifiable 
to use a non-spherical model when the single-dish observations does not contain 
any spatial information. Although the scattered-light and interferometer images
show that the structure is non-spherical, one could argue that introducing the 
flattening is simply a way to improve the fit by adding a free parameter. 
However, this is not the case. Figure~\ref{duo} shows the best fit to the 
HCO$^+$ $J =$ 4--3 line, with a spherical model ($f=0$) and a model where the 
flattening and inclination are free parameters. The flattened model provides a
considerably better fit with lower $\chi^2$ value. And, more importantly, the 
spherical model shows too much self absorption. Our optimisation algorithm 
returns the model which minimises the difference between data and model for 
each velocity channel. In a spherical model the column simply becomes too high. 
By flattening the model and adjusting the inclination, we can exactly reproduce 
the right amount of self absorption seen in the data while keeping the total 
column density high enough to produce the right line strength. The degeneracy 
between these two parameters are resolved by the velocity field, and thus it 
turns out that it is actually possible to retrieve spatial information from 
single dish observations.

\citet{ward-thompson2001} have argued that the amount of self-absorption can 
also be regulated by adjusting the turbulent velocity dispersion. We can indeed 
change the quality of the spherical fit by changing the amount of turbulence. 
We cannot, however, do that without also changing the velocity distance between 
the two peaks. Thus in order to fit the line width we must decrease the mass
and thereby the magnitude of the velocity field, which no longer reproduces the 
observed infall asymmetry. We therefore find that varying the turbulent 
velocity width in a spherical model does not reproduce the observations.

\subsection{Comparison to other observations}

With the fit parameters derived above, we can now test our model by comparison 
to other observations of L1489~IRS not used in the fit. The neighbouring cloud 
is not considered in the following.

\subsubsection{Interferometer image of HCO$^+$ 1--0}
\begin{figure}
\begin{center}
\begin{tabular}{c c}
\includegraphics[width=8.3cm, height=7.5cm]{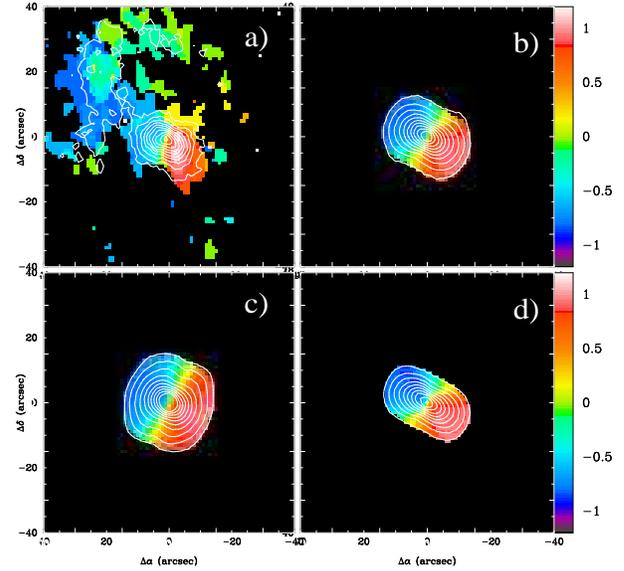}
\end{tabular}
\caption{a) Interferometer map in HCO$^+$ $J=$ 1--0. b) The corresponding 
		 synthesised map based on our best model. c) and d) Models with $f=1$ 
		 and $f=8$ respectively.  The white contour lines show the integrated 
		 intensity, starting at 0.25~Jy~bm$^{-1}$ and increasing in steps of 
		 0.5~Jy~bm$^{-1}$. The colour scale shows the velocity centroid in 
		 units of km~s$^{-1}$.
		 }\label{hcop_10}
\end{center}
\end{figure}

We have used our model to produce a synthetic interferometer map of the HCO$^+$ 
$J$=1--0 emission which can be directly compared to the data presented by 
\citet{hogerheijde2001}. Figure~\ref{hcop_10} compares the model predictions of 
the integrated intensity and velocity centroid maps to the observations 
reproduced from \citet{hogerheijde2001}. We also show a model with $f=1$ and on 
with $f=8$ for comparison. The model images were made by taking the unconvolved 
image cubes from RATRAN and then, using the $(u,v)$ settings from the original 
data set, making synthetic visibilities with the `uvmodel' task from the MIRIAD 
software package. This is what we would get if the interferometer observed our 
model object. Then we applied the usual deconvolution with the invert, clean, 
and restore routines in order to reconstruct an image from the visibilities. 

The resulting synthetic image of our best fit model resembles the observations 
closely within the uncertainty of the abundance which sensitively affects the 
apparent size, when it is taken into account that the observations also 
partially recover the neighbouring core that is ignored in the model 
(Fig.~\ref{hcop_10}, panel b). However, the $f=8$ model in panel d) is also in 
good agreement with the data which partially can be explained by the fact that, 
due to the non-linearity of the sine function, the difference between an $f=3$ 
and $f=10$ model is much less than the difference between an $f=1$ and $f=3$ 
model. This is also reflected in Fig~\ref{flatness}. In any case, panel c) is 
obviously in poor agreement with the data, which shows that in order to fit the 
interferometer data, a flattened structure is needed.

This kind of analysis is very useful to investigate the spatial distribution of 
the emission which is lacking in the single dish data. Importantly, we see that 
the flattening which we introduced and the amount of which we determined from 
the line profile results in a projected shape that is very close to what we see 
in the data image. Also, because the cuts are the same in both panels, the 
extent of the emission and therefore the physical size scale of the model is
consistent.

\subsubsection{Near-infrared image}\label{nir}

\begin{figure}
\includegraphics[width=8.5cm]{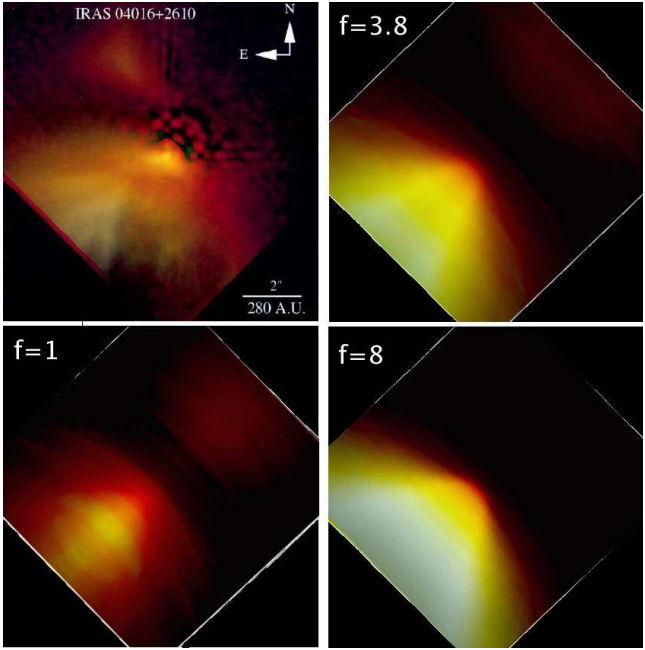}
\caption{The top left panel shows the false colour NICMOS image of L1489~IRS 
		 from~\citet{padgett1999}. In the three remaining panels are shown 
		 synthetic three-colour composite images based on our model.
		 }\label{scatter}
\end{figure}

We subsequently test our best-fit model through a comparison with the 
near-infrared scattered light image of \citet{padgett1999}. A reproduction of 
this image is shown in the top left panel of Fig.~\ref{scatter}. Using the 
density structure and inclination found in section~\ref{results} and the 
properties of the central star as described in the introduction as input, we 
calculated the scattering light emission with RADMC, a two-dimensional 
radiation transfer code \citep{dullemond2004}. Then, using the ray-tracing code
RADICAL \citep{dullemond2000}, we extract the fluxes over the full spectral 
range of 1 to 850 $\mu$m and image our model in the $F110W$, $F160W$, $F210W$ 
NICMOS bands.

Our best fit model of section~\ref{results} automatically provides a good fit 
to the sub-millimetre part of the spectral energy distribution, confirming that 
our model is consistent with the results of~\citet{jorgensen2002}. The 
near-infrared observations are more difficult to match. The column density in 
the inner part is too high to provide the clear view of the central star that 
shows prominently in NICMOS images. However, if we reduce the scale height in
the inner 250~AU to $h=0.15R$, the column density is reduced and the central 
star becomes detectable in the near-infrared. This adaptation does not affect 
the sub-millimetre emission or the molecular line intensities viewed in the 
much larger single-dish beams. Recently obtained high-resolution data from the 
Submillimeter Array (SMA) in Hawaii may be able to test this assumption of the 
geometry in the inner 2\arcsec = 280 AU (Brinch et al. in prep.).

The resulting three-colour composite images of our best fit model as well as a 
f=1 and f=8 model, is shown in Fig.~\ref{scatter}. Although this can only be a 
qualitative comparison, our model is able to reproduce most of the striking 
features evidenced by the observations. The opening angle of the dust cavity is 
found to be somewhat dependent on the flattening parameter $f$ in our model. 
Although it is difficult to judge the agreement, this result seems to favour a 
relatively low value of $f$ as opposed to what we found above for the 
interferometric map. The non-monotonic behaviour along the axis in the $f=1$ 
model is a combination of the finite gridding of the density structure and the 
effect of scattering. It has a very narrow cavity and so the base of the 
scattering nebula is actually absorbed by the high density material in the 
inner part. In the other models the cavity is large enough to allow all the 
scattered photons to escape.

We also reproduce the near-infrared colours, confirming our adopted density 
distribution between $\sim$100 AU and $\sim$2000 AU. Each of the NICMOS fluxes 
are reproduces to within about 40\%. Although a detailed modelling of L1489~IRS 
on these scales where most of the near-infrared emission is coming from was 
beyond the scope of this paper, we present this prediction to demonstrate that 
it is possible to combine the information contained in near-infrared images and 
sub-millimetre single dish measurements to obtain a self-consistent model on 
scales ranging from within a few AU up to several thousand AU.

The cavity seen in the NICMOS image is likely associated with a molecular 
outflow. However, \citet{hogerheijde1998} show that L1489~IRS only drives a 
modest $^{12}$CO outflow, and little or no impact on the line profiles is 
expected. Therefore, an outflow has not been incorporated into the model 
described in this paper.

\subsubsection{$^{12}$CO 4.7$\mu$m absorption bands}

\begin{figure}
\includegraphics[width=8.5cm]{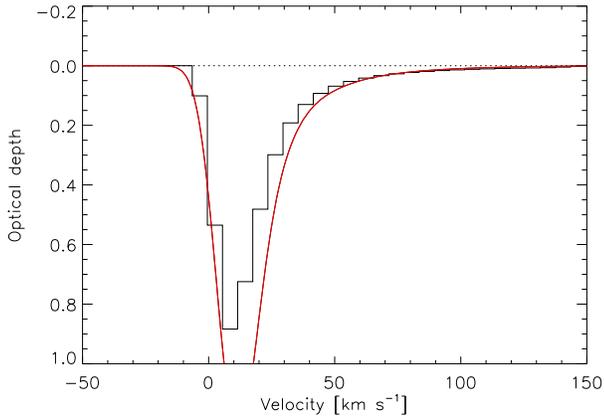}
\caption{An average of the observed $^{12}$CO P(6)-P(15) lines with a similarly 
		 averaged model result over plotted.
		 }\label{co_abs}
\end{figure}

Finally we apply our model to fit the CO ro-vibrational absorption bands, again 
by using RATRAN. Here we assume that initially all CO molecules are in the 
vibrational ground state ($v$=0) but can be excited to a $v$=1 state by 
absorption of a photon from the central star, producing the absorption lines in 
the $P$ and $R$ branches corresponding to $\Delta J=\pm1$ transitions.
Observations of these bands from the Keck/NIRSPEC instrument has been presented 
by \citet{boogert2002}. The spectrally resolved absorption lines revealed 
inward motions up to $~100$ km s$^{-1}$. Using the same method as 
\citeauthor{boogert2002} we calculated the ro-vibrational absorption lines in 
our model, and plot the average of the $P$(6)-$P$(15) lines in 
Fig.~\ref{co_abs}. 

We find that our model fits the data as well as the results from 
\citeauthor{boogert2002}, who use a contracting flared disc with power laws for 
the temperature, density, and infall velocity. This means that the amount of 
inward velocity that we obtained from the fit to the single dish lines together 
with the adopted density profile can explain the observed infall, even at radii 
much smaller than probed with the single dish lines. In Fig.~\ref{co_abs} we 
see that absorption is present all the way up to at least 100 km s$^{-1}$. In 
our model, this velocity translates into a radius of 0.02 AU. Material 
absorbing at 50 and 20 km s$^{-1}$ is located at 0.06 and 0.4 AU respectively. 

\section{Discussion}\label{discus} 

In this paper we derive an accurate model for the circumstellar material of 
L1489~IRS. We find that it is well described by a flattened structure with a 
radius of 2000~AU, in sub-Keplerian motion around a 1.35~M$_\odot$ central 
star. While the structure resembles discs found around T~Tauri stars 
\citep[e.g.]{simon2000}, its 2000~AU radius is much larger than T~Tauri discs 
(typically several hundred AU). Also, discs around T~Tauri stars are often well 
described by pure Keplerian motions except for a few cases, e.g. AB Aur, in 
which outwards non-Keplerian motions are measured~\citep{pi'etu2005,lin2006}. 
In this section we discuss the evolutionary state of L1489~IRS, and 
particularly whether it represents a unique case or if other forming stars may 
go through a similar stage. We start by deriving the life span of the current 
configuration. We then explore the relation of L1489~IRS to its neighbouring 
core. Finally, we discuss a number of open questions that only new observations 
can answer.

By integrating the trajectory of a particle at 2000 AU we find the infall time 
scale to be $2.3\times 10^4$ years. Dividing the 0.097~M$_\odot$ of the 
circumstellar material yields a mass accretion rate of $4.3\times10^{-6}$ 
M$_\odot$~yr$^{-1}$. Estimating the radius of the central star from the 
mass-radius relation~\citep{stahler1988} to be 4~R$_\odot$ results on an 
accretion luminosity of $L_{acc}$=G$M_*\dot{M}$/R$_* \approx$ 46 L$_\odot$ 
corresponding to about ten times the observed bolometric luminosity from the
YSO \citet{kenyon1993a}. This suggest that not all inspiraling material falls 
directly onto the star. The result is somewhat higher than the one found 
by~\citet{hogerheijde2001}, although the modelling approach used in that paper 
was completely different.

In our model we also assume a constant angle $\alpha$ for the direction of the 
velocity vector. In reality, this direction could vary with radius as the 
angular momentum distribution changes. For example, at smaller radii, $\alpha$ 
could be smaller as the velocity field more closely resembles pure Keplerian 
rotation. The inspiral time, and mass accretion rate, can therefore be very 
different. Only higher resolution observations can investigate this further.

Is L1489~IRS in any way special? No objects like it are reported in the 
literature. Since its life time is of the order of a few $10^4$~yr, roughly 
5--10\% of the embedded phase, more objects like it would be expected. It is 
possible that L1489~IRS was formed out of a core with unusually large angular 
momentum, which led to the formation of an untypically large disc. We cannot 
exclude that possibility, but hydrodynamical simulations are required to
investigate how much angular momentum would be required, and if turbulent cloud 
cores can contain such amounts of rotation.

An intriguing possibility is that the proximity of the neighbouring core is in 
some way related to L1489~IRS' special nature. The core is $60''$ (8400~AU) 
away in projection, and located in front of L1489~IRS, likely by a distance of 
comparable magnitude. Its systemic velocity is 0.4~km~s$^{-1}$ lower than that 
of L1489~IRS itself, indicating that L1489~IRS and the core currently are 
moving away from one another, at least in the direction along the line of 
sight. Could the neighbouring core be feeding material onto the disc of 
l1489~IRS? The velocity gradient is such that it merges smoothly with the 
velocities in the core (see Fig.~\ref{hcop_10}). This suggest that a physical
link between the core and the disc may exist. If the core feeds material onto 
the disc, this could be significant source of angular momentum, thus keeping 
the disc large. It is, however, not easy to understand why gas in the core 
would be gravitationally bound to the L1489~IRS star, because of the 
significant mass reservoir in the core itself of 2.9~M$_\odot$.

Another hypothesis is that L1489~IRS actually originated inside the core, but 
has since migrated away. Its current velocity offset of 0.4~km~s$^{-1}$ is 
sufficient to move it to its current location $60''$ away in a few times 
$10^5$~yr, the typical life time of an embedded YSO. We of course do not know 
how far its offset along the line of sight is, or what its three-dimensional 
velocity vector is like. Its line-of-sight velocity of 0.4~km~s$^{-1}$ is not 
very different from the velocity dispersion of T~Tauri stars and the turbulent 
motions in cloud complexes. One could propose that L1489~IRS' natal core was a 
turbulent, transient structure, and that, once formed, the YSO migrated with 
its gravitationally bound disc-like environment to a location outside the
surrounding cloud core, in effect `stripping' the Class~I object from most of 
its envelope. In this scenario we would now be seeing the inner, rotating 
Class~I envelope around L1489~IRS unobscured by the outer envelope. This might 
account for the different appearance of L1489~IRS. 

This hypothesis can be tested in two ways. First, `normal' Class~I objects can 
be studied at high spatial resolution and in dense gas tracers to explore if 
the inner envelope is dominated by rotation on 1000~AU scales. Second, 
hydrodynamical calculations can be used to explore on what time scales newly 
formed YSOs can migrate away from their nascent cloud core. Searching for other 
objects like L1489~IRS would also be very useful. Potential targets would 
appear compact in sub-millimetre continuum images, but show strong emission 
lines in dense gas tracers. L1489~IRS shows HCO$^+$ lines with intensities of
several K, while a few tenths of K is more typical for T~Tauri discs
\citep{thi2004}.

\section{Summary}\label{sum}
We have made a two-dimensional axi-symmetric disc-like model of the Young 
Stellar Object L1489~IRS. Line radiation transfer calculations produce 
synthetic spectra which can be directly compared to observations. We show that 
a flattened model gives a better description than a spherical one. Our model 
also reproduces millimeter interferometric imaging, near-infrared scattered 
light images, and CO ro-vibrational absorption spectroscopy.
   
We conclude further that the central star has a mass of 1.35~M$_\odot$. The 
velocity vectors make an angle of 15$^\circ$ with the azimuthal direction. The 
velocity field is dominated by rotation, while small but significant amount of 
infall are present. 

A neighouring cloud core is present next to L1489~IRS. This cloud is well 
modelled by generic dark cloud parameters and we argue that it is likely 
situated in front of L1489~IRS. We speculate that this core could be feeding 
high angular-momentum material onto the L1489~IRS disc, explaining the unusual 
size of this object. Another explanation could be that L1489~IRS is an ordinary 
Class~I object which has migrated away from its parental core, leaving it 
surrounded by only its gravitationally bound inner envelope. If this is true, 
L1489~IRS may provide valuable insight on the formation of protoplanetary 
discs.

This paper shows that it is possible to construct a global model of a Young 
Stellar Object that is able to fit observations on a wide range of spatial 
scales. Single dish line observations provide enough information to make highly 
detailed models of circumstellar structures, even on scales that are 
unresolved. However, on scales as small as 100~AU, the description may no 
longer be accurate, as evidenced by the near-infrared scattered light which 
suggest additional flattening of the disc-like structure. Recently obtained
interferometer imaging with the Submillimeter Array may provide more insight on 
the inner several hundred AU around L1489~IRS (Brinch et al.\ in prep.).

\acknowledgement
This research was supported by the European Research Training Network ``The 
Origin of Planetary Systems'' (PLANETS, contract number HPRN-CT-2002-00308). CB 
is supported by the European Commission through the FP6 - Marie Curie Early 
Stage Researcher Training programme. The research of MRH is supported by a VIDI 
grant from the Nederlandse Organisatie voor Wetenschappelijk Onderzoek. The 
research of JKJ is supported by NASA Origins Grant NAG5-13050

\bibliographystyle{aa}
\bibliography{5473.bbl}

\end{document}